\documentclass[epj]{webofc}
\usepackage[varg]{txfonts}   

\usepackage{dutchcal}
\usepackage{amsfonts}
\usepackage{amsmath}
\usepackage{subfigure}
\usepackage{graphicx}
\usepackage{hyperref}
\newcommand{\be}{\begin{equation}}
\newcommand{\ee}{\end{equation}}
\newcommand{\bea}{\begin{eqnarray}}
\newcommand{\eea}{\end{eqnarray}}
\newcommand{\nn}{\nonumber}

\begin{document}
\title{Extra Quarks and Bileptons in BSM Physics in a $331$ Model}

\subtitle{QCD@Work 2018}

\author{\firstname{Claudio} \lastname{Corian\`o}\inst{1}\fnsep\thanks{\email{claudio.coriano@le.infn.it}}, \firstname{Antonio} \lastname{Costantini}\inst{1}\fnsep\thanks{\email{antonio.costantini@le.infn.it}}}

\institute{Dipartimento di Matematica e Fisica "Ennio De Giorgi", \\ Universit\`a del Salento and INFN-Lecce, \\ Via Arnesano, 73100 Lecce, Italy}

\abstract{We describe some salient features of the $331_F$ (Frampton-Pisano-Pleitez) bilepton model, in which the constraints of anomaly cancelation require the number of generations to be three. In a class of six models, four of which characterised by a $\beta$ parameter describing the embedding of the hypercharge in the $SU(3)_L$ symmetry, a specific choice for $\beta$ allows bileptons in the spectrum, i.e. vectors and scalars of lepton numbers $\pm 2$. At the same time the model allows exotic quarks, with the third quark generation treated asymmetrically respect to the other two.
Bileptons generate specific signatures in the form of multilepton final states in Drell-Yan like processes, with and without associated jets, which can be searched for at the LHC.} 

\maketitle
\section{Introduction}
At the beginning of June 2018 the ATLAS collaboration reported the measurement of the coupling of the top quark to the Higgs boson \cite{tthATLAS}. At the same time the CMS collaboration published the results of a similar analysis \cite{tthCMS}. From their common results it is established that the coupling of the top quark to the Higgs boson is compatible with the prediction of the Standard Model (SM),  confirming that the mechanism of mass generation of the heaviest known quark is similar to that of the lighter ones \cite{hbbATLAS, hbbCMS}. These findings secure the mechanism of spontaneous symmetry breaking (SSB) - induced by a scalar sector and associated Yukawa couplings - as a general feature of the SM. If we consider BSM (beyond the SM) models, in particular models with multiple SSB steps, by the same token, it is natural to expect that extra fermionic states will be generated, widely distributed in their mass. Their charges 
are directly linked to the underlying patterns of anomaly cancellation, which, in some cases, take to electric charges which differ by those typical of the three generations of the SM, and are classified as {\em exotics}.
\section{GUT's and 331's} 
 The idea that the gauge symmetry ($SU(3)_C\times SU(2)_L\times U(1)_Y\equiv 321$) of the SM is part of a larger one has been pursued since the 70's, after that it was noticed that the unification of the gauge couplings could be achieved, provided that the underlying gauge group was {\em simple} or a product of {\em simple} factors (see \cite{Fonseca:2016xsy} for an overview). Such symmetries with multiple {\em simple} groups can be obtained by the sequential breakings of simple groups of higher rank, most notably within $E_6$ and $SO(10)$ scenarios realized at a large scale, the grand unification scale or GUT. 
 Among these, we mention the left-right symmetric model ($SU(3)_C\times SU(2)_L\times SU(2)_R\times U(1)_{B-L}$) which can be derived from the breaking of $SO(10)$. Other examples are models based on $SU(3)_c\times SU(3)_L$, later extended by an extra $U(1)_X$ in order to get the correct fermion masses. Tree level considerations on the structure of the couplings on the $331$ models suggest that the breaking $331\to 321$ could take place at the TeV scale, rather than at the usual GUT scale of $10^{13}-10^{15}$ GeV,
 which has been the motivation for studying such class of models, especially in the flavour sector ( see \cite{Buras:2012dp, Descotes-Genon:2017ptp}). There are various realization of such models since the embeddings of 321 into 331 can be quite different. 

\section{$SU(3)_c\times SU(3)_L\times U(1)_X$: exotics in the Frampton-Pisano-Pleitez model}\label{331}

The gauge structure of the $331_F$ model of
\cite{PHF,PP} is $SU(3)_c \times SU(3)_L \times 
U(1)_X$, with the fermions in the fundamental of $SU(3)_c$ arranged into triplets of $SU(3)_L$. 
The three families of quarks are treated asymmetrically with respect to the weak $SU(3)$ $(SU(3)_L)$, with the first two families given by
\begin{equation}
Q_1=\left(
\begin{array}{c}
u_L\\
d_L\\
D_L
\end{array}
\right),\quad Q_2=\left(
\begin{array}{c}
c_L\\
s_L\\
S_L
\end{array}
\right),\quad Q_{1,2}\in({\bf 3},{\bf  3}, -1/3)
\end{equation}
under $SU(3)_c \times SU(3)_L \times 
U(1)_X$,
whereas the third family is
\begin{equation}
Q_3=\left(
\begin{array}{c}
b_L\\
t_L\\
T_L
\end{array}
\right),\quad Q_3\in({\bf 3},{\bf \bar3}, 2/3). 
\end{equation}

\begin{figure}[t]
\centering
\mbox{\subfigure[]{
\includegraphics[width=0.205\textwidth]{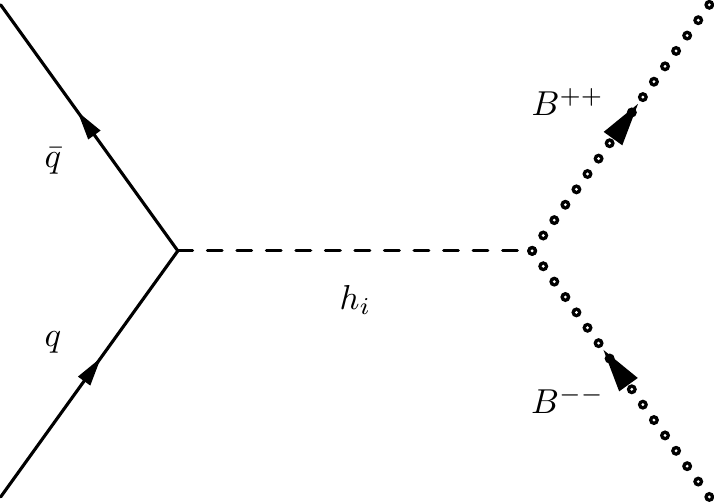}}\hspace{.8cm}
  \subfigure[]{
  \includegraphics[width=0.205\textwidth]{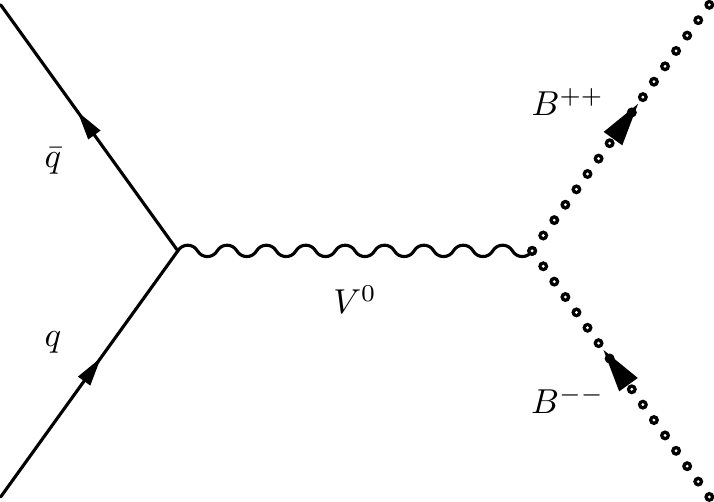}}\hspace{.8cm}
\subfigure[]{\includegraphics[width=0.205\textwidth]{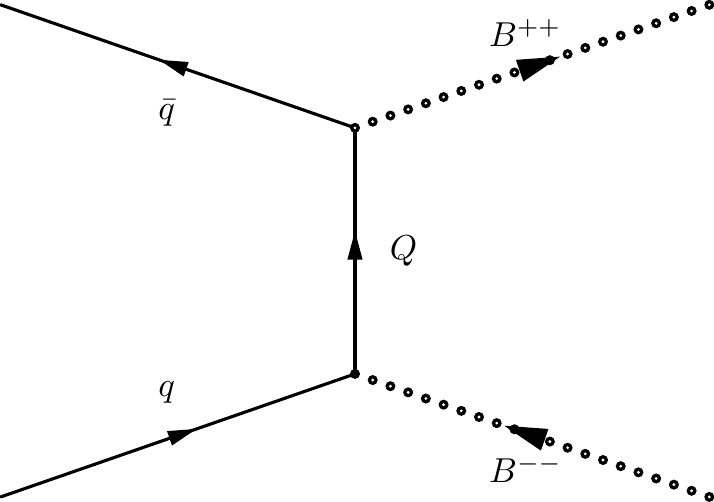}}}
\caption{Typical contributions to events with two doubly-charged bosons in the
  final state and no extra jets.
  (a) and (b): contributions due
  to the mediation of a scalar (a) and a vector (b) boson.
  (c): $t$-channel exchange of exotic quarks $Q$. $B^\pm$ denotes either a vector $(Y^\pm)$ or a scalar $(H^\pm)$ bilepton.}
 \label{jetless}
\end{figure}

$D, S$ and $T$ are extra (exotic) quarks. In this respect, the number of generations is underwritten by the contraints of anomaly cancellation, which single out this model as unique respect to the general class of BSM scenarios. The lepton sector is assigned to the representation
$\bar{3}$ of the
same gauge group. Universality is instead respected in the lepton sector, where the three lepton generations are ordinarily arranged into triplets of $SU(3)_L$, 
\begin{equation}\label{lee}
l=\left(
\begin{array}{c}
e_L\\
\nu_L\\
e_R^{\mathcal{c}}
\end{array}
\right),\quad l\in({\bf 1},{\bf \bar 3}, 0),\end{equation}
with ${e}_R^{\mathcal{c}}=i \sigma_2 e_R^*$. The presence of exotic particles in the spectrum of the 331 model
is due to the specific embedding of the 
hypercharge $Y$ in the $SU(3)_L\times U(1)_X$ gauge symmetry, which is defined by 
\begin{equation}
{Y}_{\bf 3} =\beta T_8 + X \mathbf{1} \qquad {Y}_{\bf \bar{3}} =-\beta T_8 + X \mathbf{1} 
\end{equation}
for the ${\bf 3}$ and the $\bar{\bf 3}$  representations
of $SU(3)_L$, respectively, with generators
$T_i=\lambda_i/2$ ($i=1,\ldots 8$), corresponding to the Gell-Mann matrices,
and $T_8=\textrm{diag}\left[ \frac{1}{2 \sqrt{3}}( 1,1,-2)\right]$.
The charge operator is given by 
\begin{equation} 
Q_{em, {\bf 3}}= Y_{\bf{3}} + T_3 \qquad Q_{em, \bar{\bf 3}}= Y_{\bar{\bf 3}} - T_3
\end{equation}
in the fundamental and anti-fundamental representations of 
of $SU(3)_L$, respectively, where $T_3$=${\rm diag}\left[ \frac{1}{2}
  (1,-1,0)\right]$.
 \section{Family embedding with exotics}
If we choose 
the $SU(2)_L\times U(1)_Y$ hypercharge assignments of the SM
as 
\begin{equation}
Y(Q_L)=1/6, \qquad Y(L)=-1/2,\qquad Y(u_R)=2/3,\qquad Y(d_R)=-1/3,\qquad Y(e_R)=-1
\end{equation}
and denote by $q_X$ the $U(1)_X$ charges, the breaking of the symmetry 
$SU(3)_L\times U(1)_X \to SU(2)_L\times U(1)_Y$, for the  {\bf 3} of $SU(3)$ is   
\begin{equation}
  ({\bf 3}, q_X) \to \left(2,\frac{\beta}{ 2 \sqrt{3}} +  q_X\right)
  +\left(1, -\frac{\beta}{\sqrt{3}} +  q_X\right),
\end{equation}
and for the ${\bf \bar{3}}$
\begin{equation}
  ({\bf\bar{3}}, q'_X) \to \left(2,-\frac{\beta}{2\sqrt{3}} +  q'_X\right)
  +\left(1, +\frac{\beta}{\sqrt{3}} +  q'_X\right).
\end{equation}
The $X$-charge
is fixed by the condition that the first two components of the
$Q_1$ and $Q_2$ triplets 
carry the same hypercharge 
as the quark doublets $Q_L=(u,d)_L$ of the Standard Model, yielding
\begin{equation}
\label{one}
q_X=\frac{1}{6} - \frac{\beta}{2\sqrt{3}}. 
\end{equation}

   The minimal version of the $331_F$ model, in our conventions, is obtained by $\beta=\sqrt3$, resulting in
\begin{equation}
Q_{em}(Q_{1,2})=\textrm{diag} (2/3,-1/3,-4/3), 
\end{equation}
with two exotic quarks $D$ and $S$ of charge -4/3 whereas
\begin{equation}
  Q_{em \,\bf \bar{3}}(Q_3)= \textrm{diag}\left(-\frac{1}{3},\frac{2}{3},\frac{5}{3}
  \right).
\end{equation}
The exotic quark $T$ carries electric charge 5/3. For the leptons we have the ususal charge assignments and the three generations are treated symmetrically with 
\begin{equation}
Q_{em}(l)=\textrm{diag}(-1,0,1).
\end{equation}
Both left- and right-handed components of the SM leptons are fitted into the same $SU(3)_L$ multiplet. The condition of absence of colorless fractionally charged fermions fixes the value of $\beta$ to 4 possible values $(\pm \sqrt{3}, \pm 1/\sqrt{3})$, of which the FPP model is one, plus two additional models. The first is a flipped $331$ version \cite{Fonseca:2016tbn}, with all the quarks in the same representation and the leptons in different ones; while the second is embedded in $E_6$ and there is family replication for both leptons and quarks. Among these models states with bilepton  charges are present only in $331_F$ and in the $E_6$ version \cite{Sanchez:2001ua}.

\section{The scalars of $331_{F}$}
The scalars of the $331_F$ model, responsible for the electroweak symmetry
breaking (EWSB), come in three triplets of $SU(3)_L$
\begin{equation}
\rho=\left(
\begin{array}{c}
\rho^{++}\\
\rho^+\\
\rho^0
\end{array}
\right)\in(1,3,1),\quad\eta=\left(
\begin{array}{c}
\eta^+\\
\eta^0\\
\eta^-
\end{array}
\right)\in(1,3,0),\quad\chi=\left(
\begin{array}{c}
\chi^0\\
\chi^-\\
\chi^{--}
\end{array}
\right)\in(1,3,-1).
\end{equation}

The breaking $SU(3)_L\times U(1)_X\to U(1)_{em}$ is obtained in two steps.
The vacuum expectation value  of the neutral component of $\rho$ causes
the breaking from $SU(3)_L\times U(1)_X$ to $SU(2)_L\times U(1)_Y$. Then the usual spontaneous symmetry breaking
mechanism from $SU(2)_L\times U(1)_Y$ to $U(1)_{em}$ is obtained through the vevs of the neutral components of $\eta$ and $\chi$. Exotic quarks play an important role in the production of vector bileptons $(B^\pm)$ in $q \bar{q}$ annihilaton channels, as shown in Fig. \ref{jetless}. Compared to an ordinary Drell-Yan process where the coupling of the final state leptons to the intermediate gauge bosons is direct, in this case it is mediated by the bilepton pairs, each decaying to same sign leptons with a distinct signature of multi-lepton final states. At the same time, respect to an ordinary Drell-Yan process, an extra contribution is generated by the exchange of a $t$-channel exchange of an exotic quark $(Q)$ at Born level (Fig. \ref{jetless} c). Amplitudes with no associated jets can be generated also via the sextet scalars, by replacing $Y^\pm$ with $H^\pm$.

\begin{figure}[t]
\centering\mbox{\subfigure[]{
\includegraphics[width=0.2\textwidth]{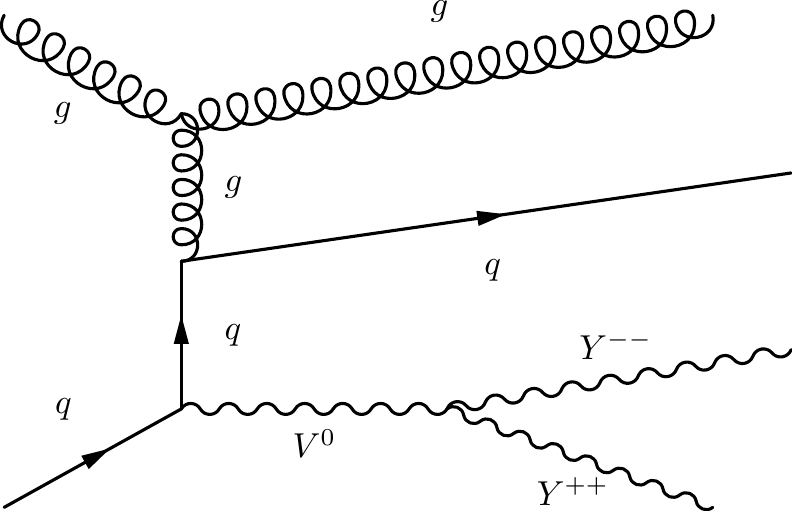}}
\subfigure[]{\includegraphics[width=0.2\textwidth]{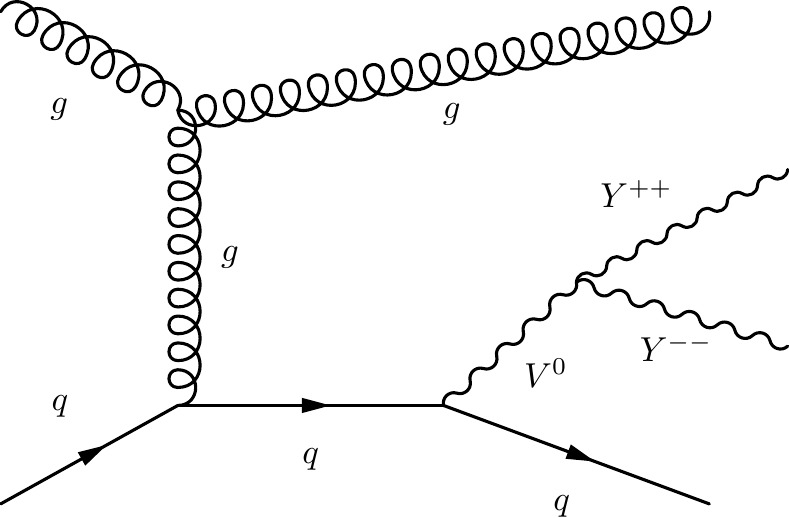}}
\subfigure[]{\includegraphics[width=0.2\textwidth]{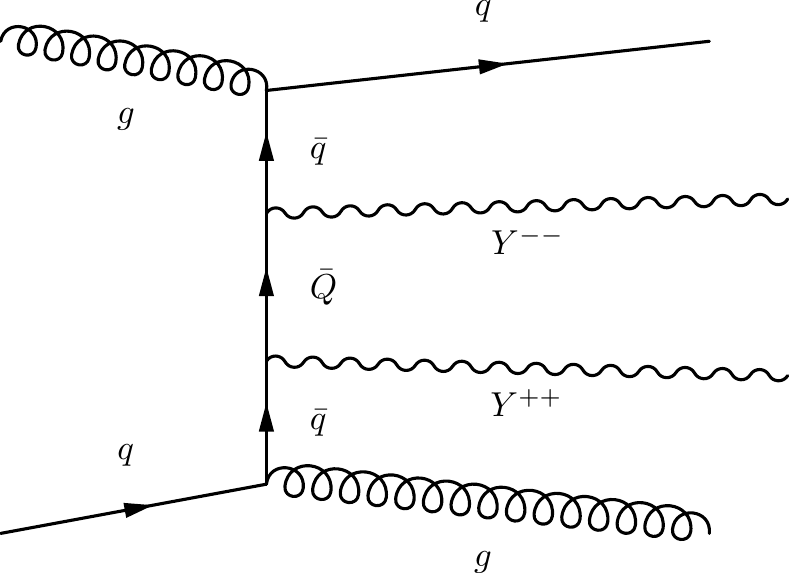}
}}
\mbox{\subfigure[]{
\includegraphics[width=0.2\textwidth]{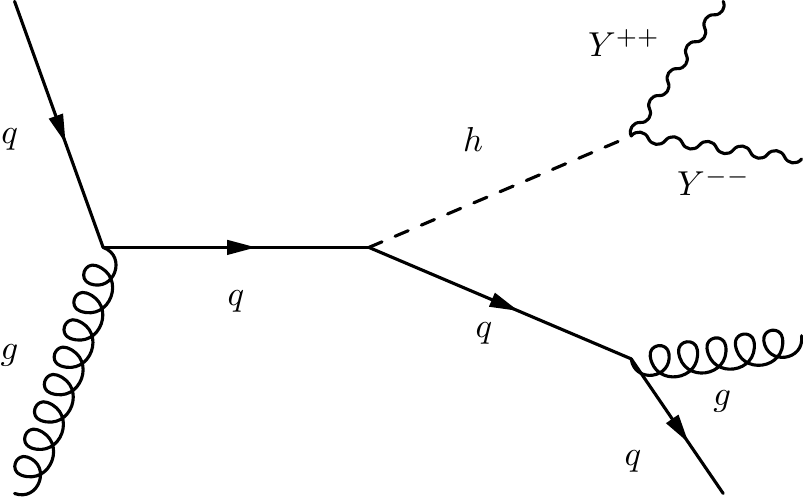}}
\subfigure[]{\includegraphics[width=0.2\textwidth]{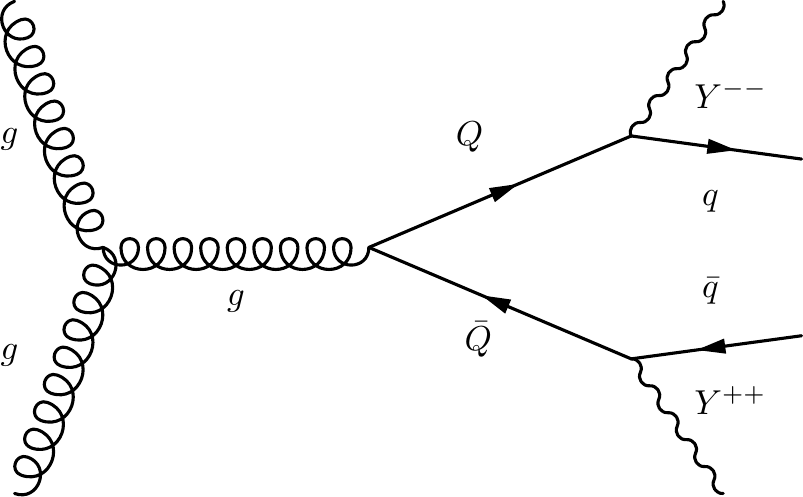}}
\subfigure[]{\includegraphics[width=0.2\textwidth]{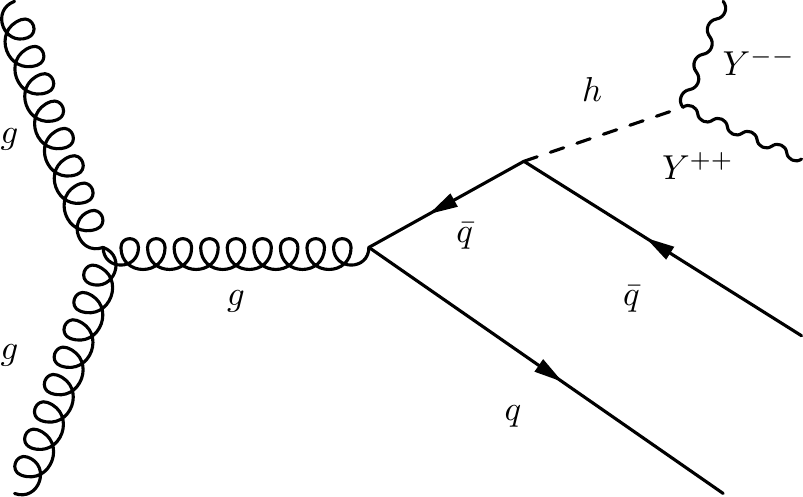}
}}
\mbox{\subfigure[]{
\includegraphics[width=0.2\textwidth]{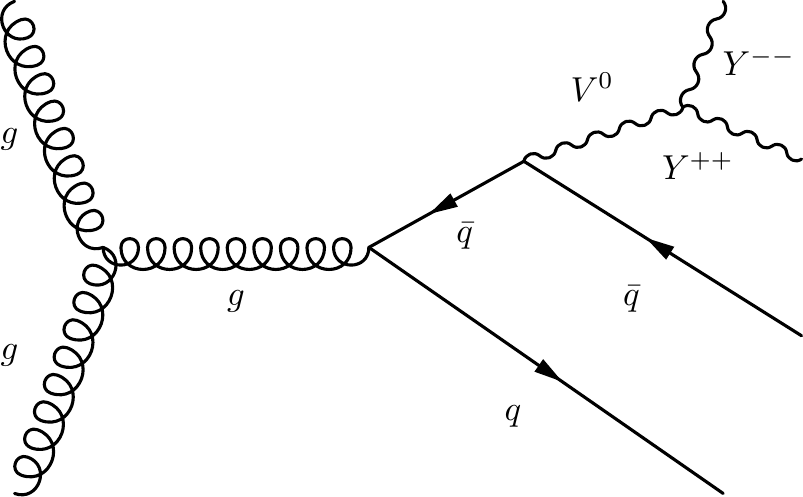}}}
\caption{Bilepton signal from quark-gluon fusion and gluon-gluon fusion.}
\label{qg1}
\end{figure}

\subsection{The triplet sector}
Coming to the electroweak  SSB mechanism, this is realised in the $331_F$ model by giving a vev to the neutral component of the triplets $\rho$, $\eta$ and $\chi$, with Yukawa interactions 
\begin{align}
\mathcal{L}_{q, triplet}^{{Yuk.}}&=\big(y_d^1\; Q_1  \eta^*  d_R + y_d^2\; Q_2  \eta^*  s_R + y_d^3\; Q_3  \chi\,  b_R^*\nn\\
&\quad + y_u^1\; Q_1  \chi^*  u_R^* + y_u^2\; Q_2  \chi^*  c_R^* + y_u^3\; Q_3  \eta\,  t_R^*\\
&\quad + y_E^1\; Q_1\,  \rho^*  D_R^* + y_E^2\; Q_2\,  \rho^*  S_R^* + y_E^3\; Q_3\,  \rho\,  T_R^*\big) + \rm{h.c.},\nn
\end{align}
where $y^i_{d}$, $y^i_u$ and $y^i_E$ are the Yukawa couplings for down-,
up-type and exotic quarks, respectively. The masses of the exotic quarks are related to the vev of the neutral component of $\rho=(0,0,v_\rho)$. In this case the invariants are, in their representation content,
\begin{eqnarray}
 Q_1\,  \rho^*  D_R^*, Q_1\,  \rho^*  S_R^*&\sim & (3,3,-1/3)\times (1,\bar{3},-1)\times (\bar{3},1,4/3) \nonumber \\
 Q_3\,  \rho  T_R^* &\sim& (3,\bar{3},2/3)\times (1,{3},1)\times (\bar{3},1,-5/3), 
\end{eqnarray}
which induce the breaking $SU(3)_c\times SU(3)_L\times U(1)_X \to SU(3)_c\times SU(2)_L\times U(1)_Y$. If the relation $y_E^i\sim1$ is satisfied, using $v_\rho\gg v_{\eta,\chi}$, it is clear that the masses of the exotics are ${\cal O}(\rm{TeV})$, 
which is of particular interest for searches of such states at the LHC. Notice that the model has the appealing feature that the condition of reality of the gauge couplings sets a constraints on the location of the extra vacua, which is limited to the TeV region.

\subsection{The sextet sector}
The triplet sector is not sufficient to account for the masses of the leptons of the SM. This is the reason for introducing a scalar sextet.
A typical Dirac mass term for the leptons in the SM
is associated with the operator $\bar{l}_LH e_R$, with $l_L=(v_{eL},e_L)$ being the $SU(2)_L $ doublet. The representation content is $(\bar{2},1/2)\times (2,1/2)\times(1,-1)$ (for $l, H$ and $e_R$, respectively) in $SU(2)_L\times U(1)_Y$. In the 331 the $L$ and $R$ components of the lepton $(e)$ are in the same multiplet and therefore the identification of an $SO(1,3)\times SU(3)_L$ singlet needs two leptons in the same representation. The flavour structure of the Yukawa terms is given by the trilinear terms 
\begin{eqnarray}
\mathcal{L}_{l,\, triplet}^{Yuk}&=& G^\eta_{a b}\, l^i_{a}\cdot l^j_{b}\,\eta^{* k}\epsilon^{i j k} + \rm{h. c.} 
\end{eqnarray}
where $a$ and $b$ are flavour indices while $i,j,k$ belong to the $3$ of $SU(3)_L$. Some considerations on the gauge charges  and fermion statistics of this interaction imply that $G^\eta_{a b}$ is antisymmetric in flavour space and as such it is not sufficient to account for the masses of all the leptons. This problem is overcome by the introduction of a scalar sector, sextet of $SU(3)_L$, denoted as $\sigma$
\begin{equation}
\sigma=\left(
\renewcommand*{\arraystretch}{1.5}
\begin{array}{ccc}
\sigma_1^{++}&\sigma_1^+/\sqrt2&\sigma^0/\sqrt2\\
\sigma_1^+/\sqrt2&\sigma_1^0&\sigma_2^-/\sqrt2\\
\sigma^0/\sqrt2&\sigma_2^-/\sqrt2&\sigma_2^{--}
\end{array}
\right)\in(1,6,0),
\end{equation}
leading to the Yukawa term
\begin{equation}\label{lag}
\mathcal{L}_{l, sextet}^{{Yuk.}}= G^\sigma_{a b} l^i_a\cdot l^j_b \sigma^*_{i,j},
\end{equation}
with $G^\sigma_{a b}$ which is symmetric in flavour space. This allows to singlet out the representation
$6$ of $SU(3)_L$, contained in $l^i_a\cdot l^j_b$, by combining it with
the flavour-symmetric $\sigma^*$, i.e. $\bar{6}$. Multi-lepton final states are a specific feature of the model generated by the decay of 
doubly charged vectors and scalars. 

\section{Quark exotics in Drell-Yan like distributions}
We show in Fig. \ref{qg1} the amplitudes involved in the analysis of bileptons with associated jets in the final state. Each exotic $(Q)$ and non exotic $(q)$ quark contributions interfere in the production of the intermediate $Y^\pm$. In Drell-Yan, as usual, the $p_T$ (transverse momentum) distributions of a lepton pair refer to processes containing 2 leptons of opposite signs plus one (quark or gluon) extra jet in the final state, at leading order in the QCD coupling. The contributions in Fig. \ref{qg1} can be viewed as a (more complex)  generalization of that process, with the total $p_T$ of the 4 leptons from the decay of $Y^\pm$ vectors balanced by those of the two extra jets. The number of amplitudes which contribute to such final states is quite large and results have been presented in \cite{Corcella:2017dns}, to which we refer for more details. Such processes can equally be generated by a pair of $H^\pm$ states if the contribution of the scalar sextet is included. \\
The inclusion of the sextet representation in the potential $V(\rho,\eta,\chi,\sigma)$ enlarges the number of physical states in the spectrum, allowing a richer scenario. In fact we now have, after electroweak SSB five scalar Higgses, three pseudoscalar Higgses, four charged Higgses and three doubly-charged Higgses.  For instance, the physical doubly-charged Higgs states are expressed in
terms of the gauge eigenstates and the elements of the
rotation matrix ${\rm R}^C$ as  \cite{Coriano:2018coq}
\bea
H_i^{++} ={\rm R}^{2C}_{i1}\rho^{++} + {\rm R}^{2C}_{i2}(\chi^{--})^* + {\rm R}^{2C}_{i3}\sigma_1^{++} + {\rm R}^{2C}_{i4}(\sigma_2^{--})^*.
\eea
In \cite{Corcella:2017dns} the study has been focused only on processes involving intermediate vectors rather than scalars, by considering benchmark points where the scalar bileptons are much heavier than the vector ones. In \cite{Coriano:2018coq}, on the other hand, the study has been focused on the Drell-Yan like contributions shown in Fig. \ref{jetless}, which provide the golden plated mode for the study of bileptons at the LHC. 
In this second study the $H^\pm$ and $Y^\pm$ states have been chosen to be close in mass. In this case the analysis has been focused on observables of the final state leptons which may allow to distinguish between processes mediated either by the $Y^\pm$ or by 
the $H^\pm$. Setting a distinction between the scalar $(H^\pm)$ and vector $(Y^\pm)$ contributions to the 4-lepton channel is an important experimental challenge, since searches for doubly-charged scalars are performed in a model-independent way, which differ from the scenario offered by this specific model, due to the presence of vector bileptons. This and the presence of exotics are peculiarities of the model which we hope will be addressed in direct experimental searches at the LHC.

\section{Conclusions}\label{concl}
In the class of  models models with a 331 symmetry, the FPP model has a special status. It carries specific signatures in the form of scalar and vector bileptons which can be searched for at the LHC. Drell-Yan like processes are the golden plated modes 
for the discovery of such realizations. We have briefly overviewed their spectra and the the specific embedding of the hypercharge in the $SU(3)_L$ symmetry which take to exotic quarks in the spectrum. The asymmetric treatment of the 3 generations is a crucial signature of many 331 models, in which the number of families is underwritten by the cancellation of the gauge anomalies. At the same time, they are constrained to lay at the TeV scale.
 \\
 \\
\centerline{\bf Acknowledgements} 
We thank Paul Frampton and Gennaro Corcella for collaborating to this analysis. This work is partially supported by INFN of Italy 
under Iniziativa Specifica QFT-HEP.

\end{document}